\documentclass[iop]{emulateapj}
\usepackage{natbib}

\shorttitle{The mid-life crisis of the Milky Way and M31.}
\shortauthors{Mutch et. al.}

\slugcomment{Accepted for publication in ApJ - May 11, 2011.}

\begin{document}

\title{The mid-life crisis of the Milky Way and M31.}

\author{Simon J. Mutch, Darren J. Croton \& Gregory B. Poole}
\affil{Centre for Astrophysics \& Supercomputing, Swinburne University of Technology, PO Box 218, Hawthorn, VIC 3122, Australia}
\email{smutch@astro.swin.edu.au}

\begin{abstract}
Upcoming next generation galactic surveys, such as GAIA and HERMES, will deliver unprecedented detail about the structure and make-up of our Galaxy, the Milky Way, and promise to radically improve our understanding of it. However, to benefit our broader knowledge of galaxy formation and evolution we first need to quantify how typical the Galaxy is with respect to other galaxies of its type. Through modeling and comparison with a large sample of galaxies drawn from the Sloan Digital Sky Survey and Galaxy Zoo, we provide tentative yet tantalizing evidence to show that both the Milky Way and nearby M31 are undergoing a critical transformation of their global properties. Both appear to possess attributes that are consistent with galaxies midway between the distinct blue and red bimodal color populations. In extragalactic surveys, such `green valley' galaxies are transition objects whose star formation typically will have all but extinguished in less than ~5 Gyrs. This finding reveals the possible future of our own galactic home, and opens a new window of opportunity to study such galactic transformations up close.
\end{abstract}


\section{Introduction}

It is a natural extension of the Copernican principle that our Galaxy should be typical when compared to other galaxies of its type.  As such, the Milky Way is commonly employed as \emph{the} template for an archetypal spiral galaxy, and is frequently used to interpret both observations and simulation data. However, despite being the most closely studied galaxy in the Universe, some of the Milky Way's most basic global properties, such as its color and star formation rate, remain relatively poorly constrained.  This is due, in large part, to our location within the disk and the difficulty in taking such measurements, especially at optical wavelengths where obscuration by dust can be particularly problematic.

Next generation galactic survey instruments, such as GAIA \citep[]{Lindegren2008} and HERMES \citep[]{Wylie-deBoer2010}, will greatly improve our knowledge of the Milky Way and its global properties. The resulting benefit to our broader knowledge of galaxy formation will, however, require us to know if the Milky Way really is a typical $\mathrm{L}^{\star}$ blue spiral galaxy, and if not, in what way(s) it differs?

Important insights into this question have already been made.  For example, it has been suggested that the probability of the Milky Way possessing two satellites of similar luminosity to the Magellanic Clouds is small ($\sim$3.5\%) \citep[]{Liu2010}. It also appears that the Milky Way is under-luminous by approximately 1$\sigma$ with respect to the main locus of the Tully-Fisher relation \citep[]{Flynn2006}.  This finding is confirmed by \citet[]{Hammer2007}, who further note that there are other Galactic properties with significant deficiencies, such as its stellar mass and disk radius.  From their results they estimate that only $\sim 7\%$ of spirals in the local Universe are similar to the Milky Way with respect to K-band magnitude, circular velocity and disk scale-length.  Interestingly, based on the same properties they find that M31 is a much better template for a typical spiral.  Using low chemical abundances in the Milky Way's outskirts as evidence, it is suggested that a likely cause for the Milky Way's disparity is an extremely quiet merger history with no accretion of objects more massive than $\sim 10^9\mathrm{M_{\sun}}$ in the last 10 Gyr.  This contrasts with the majority of other spirals, such as M31, which have had more continuous episodes of mass accretion.

Further support for an uncharacteristically quiet accretion history of the Milky Way comes from the age of the Galaxy's thin disk ($\ga$10 Gyr).  \citet[]{Stewart2008} use simulations to show that around 95\% of present day Milky Way size dark matter halos have undergone at least one merger with an object more massive than the current disk in the last 10 Gyr. Presumably such a merger would have, at a minimum, efficiently heated the thin disk, which is inconsistent with current observations.

Despite these exceptional properties there is also evidence to suggest that in many other respects the Milky Way is unremarkably normal.  Using a simulated sample of Milky Way type galaxies, \citet[]{DeRossi2009} find that the Galaxy is standard in terms of its gas fraction, luminosity weighted stellar age, and stellar mass.  They did, however, find a large dispersion in these properties, stemming from a wide range of different assembly histories. In this sense, it is important to recognize that the characterization of `typical' for any galaxy may primarily depend on the property of interest.

One global property of the Milky Way which has largely escaped scrutiny is its color index. A galaxy's color index provides an important insight into its current evolutionary state and is closely linked to its past star formation history.  In this work we compare the $B$-$V$ color index of the Milky Way and M31 with galaxies drawn from the Sloan Digital Sky Survey Galaxy Zoo morphological database \citep[]{Lintott2008}. We then confirm and interpret these results using a theoretical Milky Way galaxy catalogue constructed from a semi-analytical galaxy formation model, for which we know the full star formation and color histories of each galaxy. Using color as a probe of the evolutionary state of the Milky Way and M31, we find evidence that both galaxies may be undergoing a transition on to the red sequence.

This paper is laid out as follows. In \S\ref{sec:observations} we introduce our observational data and discuss the relevant properties of the Milky Way and M31.  In \S\ref{sec:num-models} we describe the theoretical model we use to interpret our findings.  Our main results are presented in \S\ref{sec:results} -- in particular we show that both the Milky Way and M31 would probably be classified as transition galaxies if observed from the vantage point of an extragalactic survey.  We discuss our results and highlight some of the possible physical mechanisms that may explain them in \S\ref{sec:discussion}.  Finally, a summary of our findings and conclusions are presented in \S\ref{sec:conclusions}.    

Where relevant, we adopt the following $\Lambda$CDM cosmology: $\Omega_m=0.25$, $\Omega_{\Lambda}=0.75$, $\Omega_{b}=0.045$.  All results are quoted with a Hubble constant of $h=0.7$ (where $h\equiv H_0/100\ \mathrm{kms^{-1}Mpc^{-1}}$) unless otherwise stated.  The Vega magnitude system and standard optical BVRI filters are employed throughout, with the exception of SDSS $u$-$r$ colors, which are given in the AB magnitude system.

\section{Observational Data}
\label{sec:observations}

\begin{table*}
	\begin{center}
	\begin{minipage}{114mm}
	\caption{Literature values for the relevant global properties of the Milky Way and M31.}
	\label{tab:obs}
	\begin{tabular}{l||l|l|l}
	  \multicolumn{1}{l}{}&\multicolumn{1}{l}{\textbf{Property}}&
	    \multicolumn{1}{l}{\textbf{Value}}&\multicolumn{1}{l}{\textbf{Source}}\\
	  \tableline\tableline
	  \textbf{Milky Way:}&&&\\
	  &$B$-$V$&$0.83 \pm 0.15$&\citet[]{vanderKruit1986}\\
	  &$\mathrm{M_{\star}}$&$(5.18 \pm 0.5) \times 10^{10}\ \mathrm{M_{\odot}}$&\citet[]{Flynn2006}\\
	  &SFR&$0.68 - 1.45\ \mathrm{M_{\odot}yr^{-1}}$&\citet[]{Robitaille2010}\\
	  \tableline
	  \textbf{M31:}&&&\\
	  &$B$-$V$&$0.86 \pm 0.04$\tablenotemark{a}&\citet[]{deVaucouleurs1991}\\
	  &$\mathrm{M_{\star}}$&$(10.4 \pm 0.5) \times 10^{10}\ \mathrm{M_{\odot}}$&\citet[]{Geehan2006}\\
	  &SFR&$0.41 - 0.59\ \mathrm{M_{\odot}yr^{-1}}$&\citet[]{Kang2009}\\
	  \tableline\tableline
	\end{tabular}
	\tablenotetext{1}{Includes foreground extinction correction \citep[]{Schlegel1998}.}
	\end{minipage}
	\end{center}
\end{table*}

In Table \ref{tab:obs}, we present a summary of the three key properties of the Milky Way and M31 to be studied in this paper: $B$-$V$ integrated color, stellar mass, and star formation rate (SFR). Central to our work is a comparison with a larger, cosmologically selected local `control' population, drawn from the Sloan Digital Sky Survey (SDSS) and Galaxy Zoo data sets. To enable this we have scoured the literature to find the most current values for the Milky Way and M31.

\subsection{The Milky Way}

There is much confusion and a great number of disparate results in the literature for many of the Milky Way's properties. Much of this is due to the inherit difficulties associated with observationally determining the properties of a spiral galaxy which is not seen face on and is extended on the sky. This problem is compounded for the Milky Way due to our position within the Galaxy itself.

Take, for example, the commonly quoted but little discussed $B$-$V$ color value of $B$-$V=0.83 \pm 0.15$ \citep[]{vanderKruit1986}. It was derived by comparing the Galactic background light, as observed by the Pioneer 10 probe (launched in 1972), to models of the Galaxy's stellar distribution. It is currently the most commonly cited estimate for this property, despite its age, and is found in a number of more recent publications \citep[]{Yin2009,Flynn2006}. Although there are a small number of other estimates in the literature, they are often conflicting and many are not sourced from refereed papers.  It is out with the scope of this work to discuss all of these values in detail, however, we note that, taking into account the large systematic errors associated with making such a measurement, they are all in approximate statistical agreement.

There are a similarly small number of estimates for the total stellar mass of the Milky Way. The value of $M_* = (5.18 \pm 0.33) \times 10^{10}\ \mathrm{M_{\odot}}$ comes from extrapolating local stellar surface density measurements to the full disk and adding a bulge mass estimated using dynamical constraints \citep[]{Flynn2006}. The quoted uncertainty results from considering a realistic range of assumed disk scale lengths. The true uncertainty on this value is likely considerably larger, however, with no real understanding of the underlying systematics we choose to simply round the quoted uncertainty up to $\pm 0.5 \times 10^{10} \mathrm{M_{\odot}}$ for this work. Similar stellar mass values for the Milky Way are obtained by converting \textit{K}-band measurements to a stellar mass value via color dependent $\mathrm{M}_{\odot}/L_{K}$ ratios \citep[]{Hammer2007}. It should be noted, however, that a slightly bluer color value for the Galaxy of $B$-$V$=0.79 has been utilized during the derivation of this result (note that this is still in good statistical agreement with our fiducial value of $B$-$V=0.83 \pm 0.15$).

The most up-to-date determination of the Milky Way's current star formation rate is $\dot{M}_* = 0.68 - 1.45\ \mathrm{M_{\odot}yr^{-1}}$ \citep[]{Robitaille2010}. This is calculated by comparing the number of Spitzer/IRAC GLIMPSE observed pre-main-sequence young stellar objects to population synthesis models. The derived value is slightly lower than, but still broadly consistent with, previous estimates \citep[e.g.]{Rana1991,Boissier1999,Fraternali2009}. However, \citet[]{Robitaille2010} argue that this difference is largely due to the assumed IMF, with their choice being more realistic, especially at low stellar masses. This value is also directly derived from actual statistics of pre-main sequence objects rather than simply from global observables.

Finally, we assume the Hubble type of the Milky Way to be approximately Sb/c \citep{Kennicutt2001,Hodge1983}. 

\subsection{M31}

Due to the benefit of its large angular size on the sky, the majority of authors spatially decompose the properties of M31 rather than use its integrated properties. The best global $B$-$V$ color estimate of $0.92\pm0.02$ comes from the Third Reference Catalogue of Bright Galaxies (RC3) \citep[]{deVaucouleurs1991}. This value has not been corrected for either internal or Galactic extinction and so we apply a foreground extinction correction of E($B$-$V$)=0.062 \citep[]{Schlegel1998} to arrive at the final color listed in Table ~\ref{tab:obs}: $B$-$V = 0.86 \pm 0.04$. We note that the RC3 catalogue value is in excellent agreement with a number of independent works \citep[]{deVaucouleurs1958,Walterbos1987} and hence we are confident in its fidelity.

Our utilized measurement for the global stellar mass of M31 was determined through the summation of the disk and bulge component from a semi-analytic mass model \citep[]{Geehan2006}. The resulting value of $M_{\star} = 10.4\times 10^{10}\ \mathrm{M_{\odot}}$ is consistent with the findings of a number of other recent works \citep[e.g.][]{Hammer2007,Barmby2006}.  Considering the range of values quoted in the literature, we choose to estimate an uncertainty of $\pm 0.5 \times 10^{10} \mathrm{M_{\odot}}$.

The star formation rate of M31 is much better established than that of the Milky Way. The range of allowed values is $\dot{M}_* = 0.41 - 0.59\ \mathrm{M_{\odot}yr^{-1}}$ and is derived from combined UV and IR fluxes in star formation regions with an age of less than approximately $0.4 \mathrm{Myr}$ \citep[]{Kang2009}. However, we note that this range of SFR values is identical, to within $\sim$0.1 $\mathrm{M_{\odot}yr^{-1}}$, to the value when averaged over the last 400 $\mathrm{Myr}$. The mid point quoted in Table~\ref{tab:obs} is for an assumed metallicity of $\mathrm{Z}=0.02$, with the upper and lower boundaries corresponding to $\mathrm{Z}=0.05$ and $\mathrm{Z}=0.008$ respectively.

Following \citet[]{Tully2000}, we take the Hubble type of M31 to be approximately Sb.

\subsection{Galaxy Zoo Control Sample}
\label{sec:galaxy-zoo}

To fairly interpret the properties of the Milky Way and M31 we employ a large control sample of $z\approx 0$ galaxies from \citet[]{Bamford2009}; a luminosity-limited catalogue of $\sim 130\,000$ galaxies taken from Sloan Digital Sky Survey \citep[SDSS (DR6);][]{Adelman-McCarthy2008} data and cross correlated with Galaxy Zoo1 \citep[]{Lintott2011} visual morphologies.  The sample has been selected to contain galaxies with spectroscopic redshifts in the range $0.01<z<0.085$, corresponding to an r-band absolute Petrosian magnitude limit of $M_r < -20.17$, and is carefully sub-sampled in an effort to remove any redshift dependent selection biases.  All magnitudes are in the AB system and are k-corrected to $z=0$.  Stellar masses are derived using the method of fitted color dependent mass-to-light ratios discussed in \citet[]{Baldry2006}.  We refer the reader to \citet[]{Bamford2009} for further details.  We henceforth refer to this sample as the \emph{full Galaxy Zoo sample}.

In order to produce a sample of Milky Way/M31 analogue galaxies we select only galaxies which:
\begin{itemize}
	\item are visually classified spirals with de-biased morphological likelihoods $p_{spiral}\ge 0.8$ \citep[c.f.][]{Bamford2009},
	\item have stellar masses in the range \mbox{$10.66 < \log_{10}(\mathrm{M_\star[M_{\odot}]}) < 11.12$}\\ (c.f.  $\log_{10}(M_{\star_{\mathrm{MW}}}/M_{\odot})$=10.71; $\log_{10}(M_{\star_{\mathrm{M31}}}/M_{\odot})$ = 11.02),
	\item possess an SDSS {\it fracdev} value in the range 0.1--0.5 \citep[i.e. have an approximate Sb/c morphology, c.f.][]{Masters2010},
    \item and are face-on with $log_{10}(a/b) < 0.2$, where $a/b$ is the major to minor axis ratio as observed on the sky. 
\end{itemize}
The {\it fracdev} parameter measures the relative fraction of the best fitting light profile that comes from a de Vaucouleurs fit.  Following \citet[]{Masters2010}, we use this value to approximate a galaxy's bulge-total mass ratio and hence its morphology.  This selection will, in reality, encompass a range of morphological types from $\sim$Sb-Sc.  We place no direct constraints on the visibility of spiral arms in our sample.  By selecting only face-on galaxies, we minimize, as much as possible, the systematic reddening effects of intrinsic interstellar absorption which could cause otherwise blue Sb/c galaxies to masquerade as green valley or red sequence members.  We also note that, although we have chosen to utilize a single comparison set for both the Milky Way and M31, we have confirmed that splitting our selection criteria into two equal stellar mass ranges makes no qualitative difference to our findings. 

Our resulting \emph{Galaxy Zoo Milky Way/M31 analogue sample} contains 997 objects.

\section{Modeled Data}
\label{sec:num-models}

The observational data presented in \S\ref{sec:observations} provides a reasonably precise snapshot of the current state of the Milky Way and M31, as well as a census of the broader local galaxy population. However, it does not give us any insight into the range of evolutionary histories that may have led to this final state, or of the physical influences that were pertinent. To obtain this we employ a model of galaxy evolution that explicitly tracks the full history of $L^*$ galaxy growth and the physics that shapes it. 

\subsection{A `Semi-analytic' Model of Galaxy Formation}
\label{sec:SAM}

The semi-analytic method of modeling galaxy formation was first introduced by \citet[]{White1991} and follows a two-staged approach:

\begin{itemize}
\item First, a numerical simulation of the growth of the dark matter structure of the Universe is run to establish the potential formation sites and gravitational evolution of galaxies, galaxy groups and clusters. For our work we use the Millennium Simulation \citep[]{Springel2005}, a cosmological simulation that evolves $\sim$10 billion dark matter particles in a $1.24 \times 10^8~\rm{Mpc}^3 h^{-3}$ representative volume. The Millennium Simulation accurately tracks the hierarchical growth of structure from two orders-of-magnitude smaller than a Milky Way system (i.e. dwarfs), to three orders-of-magnitude larger (i.e. clusters).
\item Second, analytic prescriptions that describe the physics of galaxy formation are coupled to the dark matter simulation to predict the evolution of baryons in the evolving dark matter structure. For this work we use the commonly utilized model of \citet[]{Croton2006}. This model is calibrated to produce a good global match to the local galaxy population in terms of its stellar mass and luminosity functions, galaxy colors, morphologies, and clustering \citep[]{DeLucia2007, Croton2007, Kitzbichler2008}. To achieve this the model employs prescriptions that describe gas accretion and cooling in dark matter halos, galaxy disk formation, star formation and dust, supernova feedback, the production and evolution of metals, galaxy-galaxy mergers and morphological transformations, black hole growth and active galactic nuclei feedback. 
\end{itemize}

See \citet[]{Croton2006} for a more detailed description of each component of the model and its construction.

\subsection{Selecting the Theoretical Milky Way Analogue Sample}
\label{sec:MW-sample}

Once the model is run we have on hand a large number of simulated galaxies at $z=0$ ($\sim$25 million) with well defined properties (masses, luminosities, colors, SFRs, ...) and complete evolutionary histories. From these galaxies we sub-select Milky Way/M31 analogues to be used to interpret the observed state of the Milky Way and M31.

Following a set of criteria which are as similar as possible to that of the Galaxy Zoo sample (\S\ref{sec:galaxy-zoo}), we select Milky Way/M31 analogues as all those galaxies from the model with the following well defined properties\footnote{Note that we include the effects of dust extinction on galaxy luminosity through a simple `plane-parallel slab' model \citep[]{Kauffmann1999}, and assume a Chabrier initial mass function when calculating galaxy masses \citep[]{Chabrier2003}.}.  They must:
\begin{itemize} 
    \item be the most massive (i.e. central) galaxy of their dark matter halo, 
    \item possess a stellar mass in the range \mbox{$10.66<\log_{10}(\mathrm{M_\star[M_{\sun}]})<11.12$} (identical to our face-on Galaxy Zoo Milky Way/M31 analogue sample), 
    \item possess an approximate Sb/c morphology with a bulge-total luminosity ratio of $1.5<M^{bulge}_{B}-M^{total}_{B}<2.6$ \citep[]{Simien1986}, and 
    \item be face-on with $\log_{10}(a/b) < 0.2$ (reducing the effects of internal dust extinction and again mimicking our face-on Galaxy Zoo Milky Way/M31 analogue sample criteria). 
\end{itemize}
These criteria result in a sample of $28\,439$ Milky Way-type galaxies which we refer to as the \emph{theoretical Milky Way/M31 analogue sample}.  Note the larger number of galaxies in this sample when compared to the Galaxy Zoo analogue sample.  This is predominantly due to the increased volume probed by the simulation.  We also note that the use of the above bulge-to-total luminosity ratio to define galaxy morphology is again an approximation which will result in the selection of galaxies of $\sim$Sb-Sc morphology.

\section{Results}
\label{sec:results}

\subsection{The Position of the Milky Way and M31 on the Color--Stellar Mass Diagram}
\label{sec:obs-results}

\begin{figure*}
	\begin{center}
	\begin{minipage}{160mm}
    \includegraphics[width=160mm]{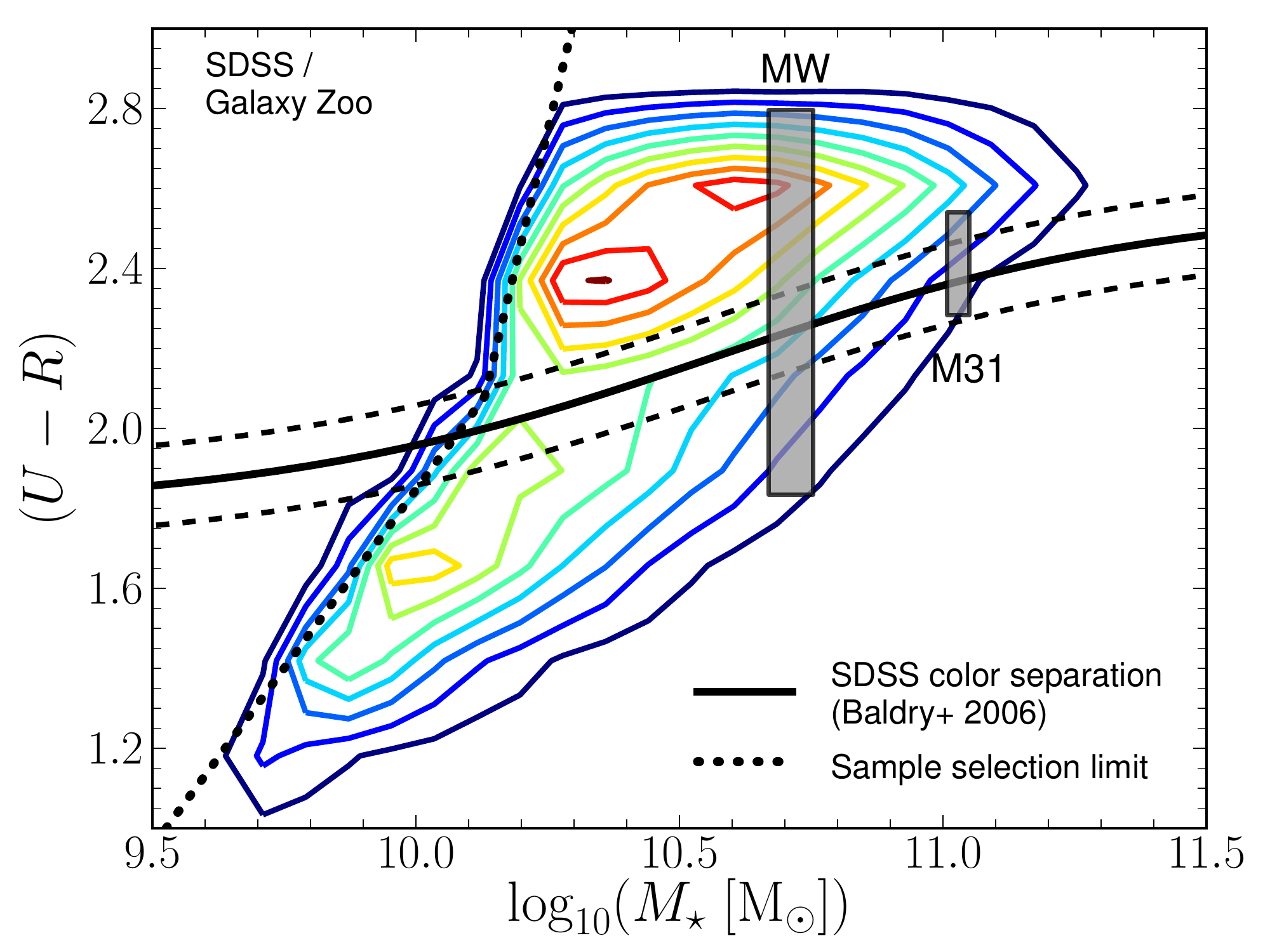}
    \caption{\label{fig:colourSM_SDSS_contours} Contour plot of $u$-$r$ color vs. stellar mass for the full Galaxy Zoo sample. The contours are linearly spaced. The stellar mass corresponding to the absolute magnitude limit of the sample is plotted as a dotted line. The solid line indicates the location of an independently derived best fit bimodal color population division \citep[]{Baldry2006}. The bounding dashed lines are $\pm$0.1 $u$-$r$ of this division and delineate our definition of the green valley region. The locations of the Milky Way and M31 are indicated using the values listed in Table \ref{tab:obs} after converting the $B$-$V$ color values to $u$-$r$ (see \S\ref{sec:obs-results}). It is apparent that M31 is a candidate green valley member, however, the large uncertainty on the color of the Milky Way precludes us from making a similar statement in its case. See the electronic edition of the Journal for a color version of this figure.}
\end{minipage}
\end{center}
\end{figure*}

Galaxy color is commonly used to identify galaxies at broadly different evolutionary stages.  As a galaxy ages, its cold gas reserves -- the raw material for star formation -- are depleted, causing star formation to subside and the primary stellar population to age and redden.  As a result, and in the absence of complications due to dust obscuration, we find that red galaxies are older and no longer forming stars, while blue galaxies are younger with active star formation and larger reserves of cold gas.

It has been well established for some time that these blue and red galaxies form distinct populations on a color vs. magnitude (and hence color vs. stellar mass) diagram \citep[]{Strateva2001, Baldry2006}.  They are commonly referred to as the `red sequence' and `blue cloud', with the sparsely occupied region between them dubbed the `green valley'.  

Figure~\ref{fig:colourSM_SDSS_contours} shows the full Galaxy Zoo observational sample (\S\ref{sec:galaxy-zoo}) plotted in the $u$-$r$ color magnitude vs. stellar mass plane with linearly spaced contours.  The dotted lines indicate the minimum stellar mass sample selection limit as a function of color, corresponding to the magnitude limit of the sample.  

Both the red sequence and blue cloud are clearly visible in Figure~\ref{fig:colourSM_SDSS_contours}, and highlight the green valley as an under-densely populated region between the two. We have overlaid the independently derived best fit bimodal color population division of \citet[]{Baldry2006} (solid curved line).  The green valley region is defined to be $\pm$0.1 $u$-$r$ of this division (bounded by dashed curved lines).  We have checked that making minor variations to assumed width of the valley region results in no qualitative change to the conclusions of this work.

We frame the Milky Way and M31 against the backdrop of the large Zoo sample by over-plotting their colors in Figure~\ref{fig:colourSM_SDSS_contours}, after converting from $B$-$V$ (Table~\ref{tab:obs}) to $u$-$r$ (SDSS model) using the following formula:
\begin{eqnarray}
 \label{eq:colour-conversion}
  (u-r) = (1/0.3116)\ \left[ (B-V) - 0.1085 \right] ~.
\end{eqnarray}
This was derived from the magnitude system conversions provided by Robert Lupton and published on the SDSS website\footnote{http://www.sdss.org/dr7/algorithms/sdssUBVRITransform.html}.

The stellar mass and resulting $u$-$r$ color of M31 is consistent with the green valley region of the full Zoo sample. However, the size of the uncertainty on the color of the Milky Way is too large to allow us to place any reasonable constraints on its color classification.  Further evidence is required to determine how statistically plausible it is that these two galaxies are indeed `green'.

\subsection{How common are green valley Milky~Way/M31-type galaxies?}
\label{sec:stats}

\begin{figure*}
	\begin{center}
	\begin{minipage}{160mm}
    \includegraphics[width=160mm]{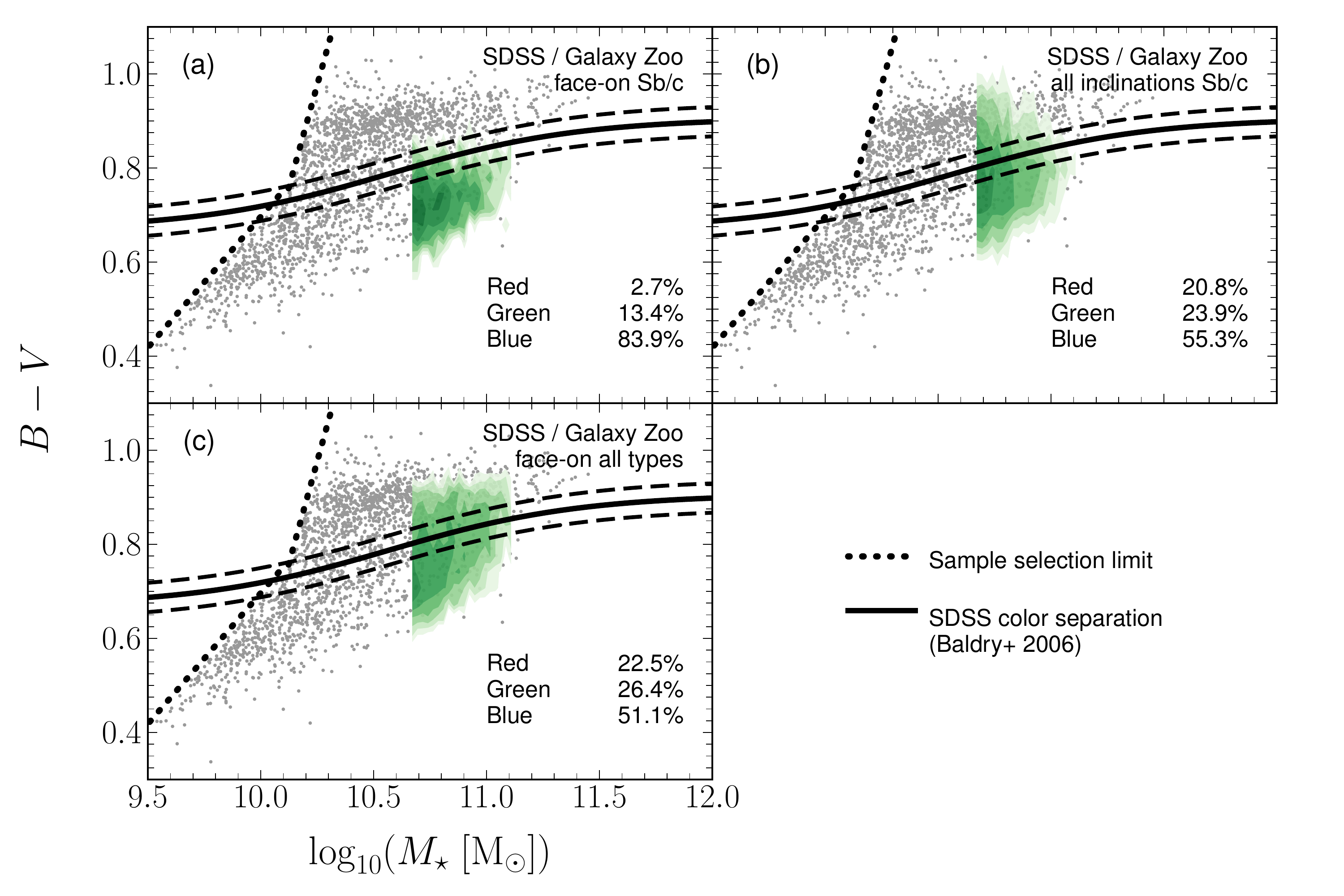}
    \caption{\label{fig:CSMD_MultiPanel} Plot of $B$-$V$ color as a function stellar mass for the Galaxy Zoo sample. Grey points represent a random sampling of the full sample population. Solid and dashed lines are as in Figure~\ref{fig:colourSM_SDSS_contours}. Green shaded contours are logarithmically spaced and show the distribution of Milky Way/M31 analogue galaxies in color space for varying selection criteria. Panel (a) shows distribution for our fiducial criteria, described in \S\ref{sec:galaxy-zoo}. Panel (b) shows how the relative color fractions change if we include galaxies of all inclinations. Finally, panel (c) demonstrates the effects of relaxing our criteria to include all spiral types (but maintaining the inclusion of only face-on galaxies). See the electronic edition of the Journal for a color version of this figure.}
\end{minipage}
\end{center}
\end{figure*}

We now look for evidence to support the location of the Milky Way and M31 in the green valley, as shown in Figure~\ref{fig:colourSM_SDSS_contours}. The Galaxy Zoo data set provides us with a statistically significant sample of Milky Way-like galaxies with measured morphologies, stellar masses and colors (\S\ref{sec:galaxy-zoo}). With it we can investigate the statistical likelihood of Sb/c spiral galaxies residing in the green valley or on the red sequence. 

Figure~\ref{fig:CSMD_MultiPanel}(a) shows the $B$-$V$ color vs. stellar mass distribution of the Galaxy Zoo Milky Way/M31 analogue sample (green contours) superimposed against the full Zoo sample used previously in Figure~\ref{fig:colourSM_SDSS_contours} (randomly sampled 2\%; grey points).  The Johnson $B$-$V$ color values were converted from the SDSS AB model magnitude $u$-$r$ colors using the inverse of the transformation provided in Eqn.~\ref{eq:colour-conversion}. The thick solid line again indicates the optimum red/blue color division derived by \citet[]{Baldry2006}.  The green valley division lines of Figure~\ref{fig:colourSM_SDSS_contours} (again converted to $B$-$V$ using Eqn.~\ref{eq:colour-conversion}) have also been over-plotted.

Our color classification results in 3\% of our Milky Way/M31 analogues being classified as red, 13\% as green and 84\% as blue.  In other words, from our Galaxy Zoo data we find that approximately 1 in 6 local Milky Way-like galaxies lie red-ward of the blue cloud.  This suggests that it is statistically plausible for both the Milky Way and M31 to be candidate green valley members.

Figure~\ref{fig:CSMD_MultiPanel}(b) is identical to Figure~\ref{fig:CSMD_MultiPanel}(a), however, here we place no restriction that galaxies must be face-on. By removing this constraint we find a considerably higher fraction of galaxies ($\sim$45\%) lying redward of the blue cloud. This indicates that many of the galaxies in the Galaxy Zoo Milky Way/M31 analogue sample likely suffer from significant reddening due to internal dust extinction and therefore do not have truly passive stellar populations. By sub-selecting only face-on galaxies in our analysis, we have reduced this effect considerably, allowing us to more realistically estimate the relative fractions of galaxies redder than the blue cloud \citep[]{Masters2010b}. However, we do note that it is still possible for the colors of our face-on sample to be contaminated by dust to some non-negligible extent \citep[]{Driver2007}. Although we make no attempt to quantify the magnitude of this effect, a simple investigation using a sample of face-on Galaxy Zoo spirals was carried out by \citet[]{Masters2010}.  They found little evidence in their sample to suggest that red face-on spirals have a significantly larger dust content, and hence suffer from a greater reddening effect, than their blue counterparts at a given stellar mass.  They hence concluded that dust reddening in their face-on sample was not a major cause for creating a red spiral population.

Figure~\ref{fig:CSMD_MultiPanel}(c) displays the distribution of galaxies from the \emph{face-on} Galaxy Zoo Milky Way/M31 analogue sample, but this time selecting all spirals, regardless of the morphology estimated by their \emph{fracdev} values. Again, we find an increased fraction of galaxies redward of the blue cloud ($\sim$50\%) when compared to our fiducial Sb/c morphology sample (Figure~\ref{fig:CSMD_MultiPanel}(a)). As all of the galaxies in this plot are still face-on, this increased fraction is likely \emph{not} an effect of dust contamination but is instead due to the inclusion of more bulge dominated spirals. This quantitative finding agrees with other works \citep[e.g.]{Bundy2010,Masters2010b} which find that red passive spirals tend to posses larger bulges than their star forming blue counterparts.

The high fraction of reddened Milky Way-type galaxies found when we relax our selection criteria demonstrates the conservative nature of our estimate that at least 1 in 6 local Milky Way/M31 analogues lie in the green valley or on the red sequence (dependent on how tightly we constrain the sample classification).

\subsection{Further constraints using the model analogue sample}
\label{sec:model-results}

\begin{figure}
    \includegraphics[width=85mm]{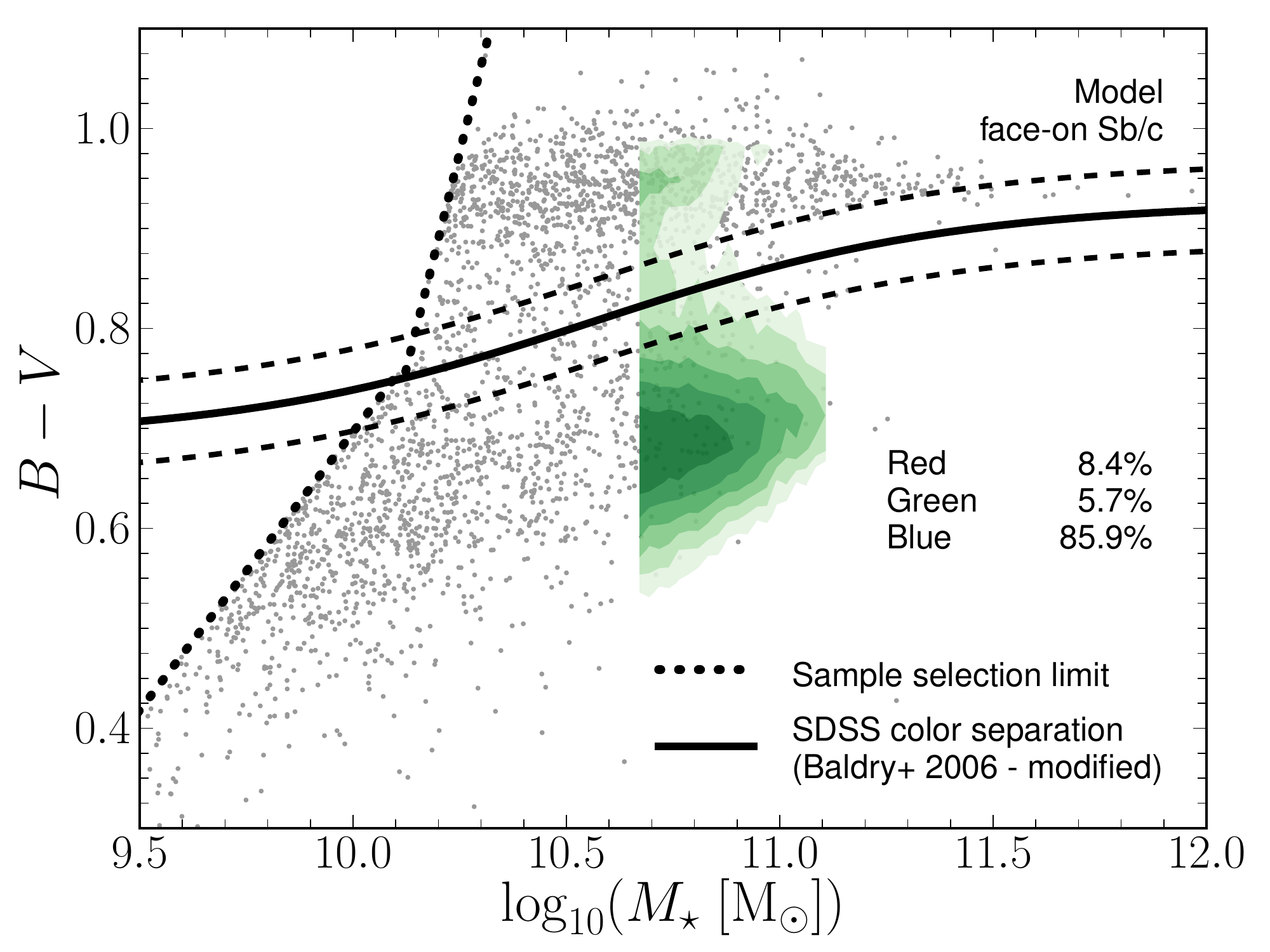}
    \caption{\label{fig:CSMD-model} $B$-$V$ color as a function stellar mass for the theoretical model galaxies. Grey points represent a random sampling of the full model output at $z$=0. Green shaded contours are logarithmically spaced and show the distribution of face-on model Milky Way/M31 analogue galaxies in color space. The solid and dashed curved lines are as described in Figure~\ref{fig:colourSM_SDSS_contours}. The green valley region has been raised and widened from that of the observational sample (Figs.~\ref{fig:colourSM_SDSS_contours}\ref{fig:CSMD_MultiPanel}) by 0.01 B-V in order to more accurately reflect the location of the green valley in the model data. Our qualitative results are, however, insensitive to the precise magnitude of these changes. There is good qualitative and quantitative agreement between the model Milky Way analogue data set, shown here, and the observational Galaxy Zoo sample, shown in Figure~\ref{fig:CSMD_MultiPanel}. See the electronic edition of the Journal for a color version of this figure.}
\end{figure}

In Figure~\ref{fig:CSMD-model}, we show an equivalent color vs. stellar mass plot for our theoretical Milky Way/M31 analogue sample at $z=0$, constructed from the galaxy formation model described in \S\ref{sec:SAM}.  In this case, the grey points represent a randomly sampled 0.03\% of the full model output and the green contours represent our model Milky Way/M31 analogues (\S\ref{sec:MW-sample}).  The green valley region has been raised by 0.01 $B$-$V$, and similarly widened by the same amount, with respect to our observational valley region so as to more accurately reflect the location of the green valley in the model data.  We confirm that making sensible variations to both the offset and width of the defined green valley region results in no difference to our qualitative conclusions.  Although the precise distribution of galaxies is not identical, the same color bimodality present in the observational data (Figure~\ref{fig:colourSM_SDSS_contours}) is clearly visible.  We find 24\,422 blue, 1\,623 green and 2\,394 red model Milky Way-like galaxies, corresponding to 86\%, 6\% and 8\% of the sample respectively.  Hence 14\% of model Milky Way galaxies lie red-ward of the blue cloud (c.f. 16\% in the observational Zoo sample (\S\ref{sec:stats})).

The general agreement between the model and Galaxy Zoo data, for both the color--stellar mass distribution and, in particular, the fraction of Sb/c spirals as a function of color, gives us confidence that our simulation is a reasonable tool with which to extend our analysis. We now look for additional correlations using stellar mass and star formation rate as a function of galaxy color, and then compare these to the observed masses and star formation rates of the Milky Way and M31.

\begin{figure}
    \includegraphics[width=85mm]{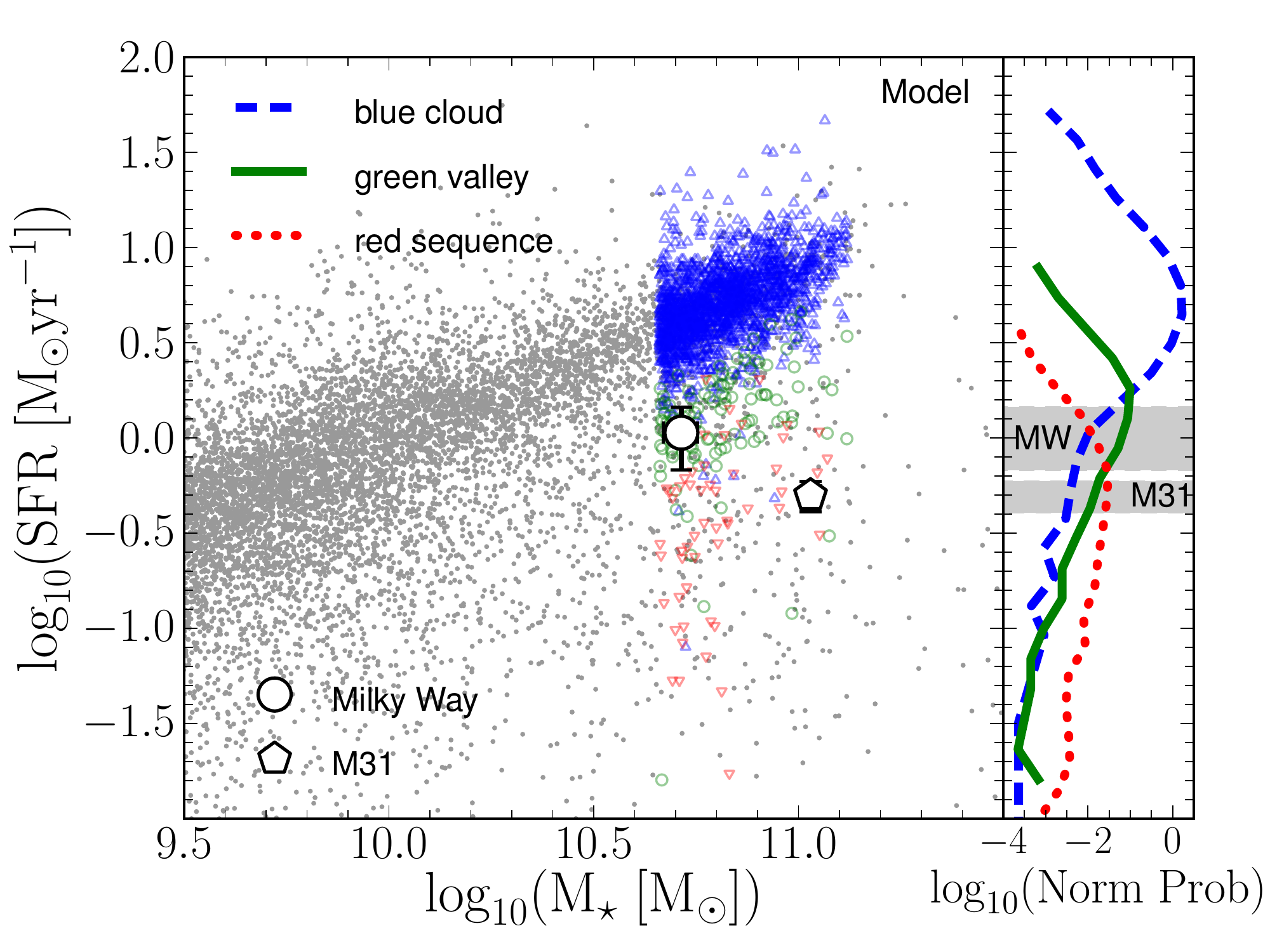}
    \caption{\label{fig:SFR-stellar} [\emph{Left}] Plot of star formation rate (SFR) as a function stellar mass for our \emph{model} galaxies. Grey points represent a random sampling of the full model output. Blue upward triangles, green circles and red downward triangles show the distribution of face-on Milky Way/M31 analogue galaxies in this simulated sample.  The real, observed locations of both the Milky Way and M31 in this parameter space are indicated using the values listed in Table \ref{tab:obs}.  The uncertainties in the stellar masses of both of these galaxies are less than, or equal to, the width of the symbols and so they have been omitted for clarity.
    [\emph{Right}] Normalized probability distribution of SFR for each Milky Way/M31 analogue color sub-sample.  The normalization is such that the probability of belonging to each color sample at any given SFR may be directly compared. The SFRs of the Milky Way and M31 are shown as horizontal shaded regions. Against the backdrop of the galaxy formation model, they are approximately 2--5 and 3--4 times more likely to be associated with the green valley than the blue cloud respectively. See the electronic edition of the Journal for a color version of this figure.}
\end{figure}

In the left-hand panel of Figure~\ref{fig:SFR-stellar} we plot the star formation rate of our model galaxies against their stellar mass.  The grey points are again a randomly sampled selection of the entire model output.  These points form a locus which displays a trend of increasing SFR with increasing stellar mass and which can largely be associated with the blue cloud galaxies on the color vs. stellar mass plot of Figure~\ref{fig:CSMD-model} \citep[]{Noeske2007}.  The red, green and blue points show the location of our theoretical Milky Way/M31 analogue sample in this space, broken down by color as classified previously in Figure~\ref{fig:CSMD-model}.  As expected, we find that blue galaxies have the highest SFRs, and red galaxies the lowest.  The actual observed positions of the Milky Way and M31 have been over-plotted using the values listed in Table \ref{tab:obs}.  Both of these galaxies have star formation rates which place them below the main locus of the model star forming galaxies.

The right-hand panel of Figure~\ref{fig:SFR-stellar} shows the normalized SFR probability distribution for our theoretical Milky Way/M31 analogue sample, split by color.  The observed SFR values of both the Milky Way and M31 are overlaid as grey shaded regions.  When compared with the model SFR distributions, the observed SFR of both the Milky Way and M31 support their previous classification as green valley members in the color--stellar mass diagram.  In the case of the Milky Way, we find the probability of it being associated with the green valley, based on its SFR, to be $\sim$ 3--5 times greater than the probability of it being associated with the blue cloud (relative to the model population).  Similarly, the SFR of M31 makes it $\sim$ 3--4 times more likely to be associated with the green valley. In fact, by this diagnostic, M31 is more likely to be red rather than green or blue.

\section{Discussion: What makes a `green' Milky Way?}
\label{sec:discussion}

Gathering together the main results of \S\ref{sec:results} suggests that both the Milky Way and M31 are possible green valley members.  In the generalized framework of galaxy evolution such objects are thought to represent those galaxies which are in the process of maturing from blue and star forming to `red and dead.'  How long does this transitional phase last?  What is the mechanism which triggers a galaxy's departure from the blue cloud?  What is the significance of a `green Milky Way'?  Our theoretical Milky Way/M31 analogue sample provides us with a powerful tool with which to gain important insights into these complex questions.

\subsection{Are Red Milky Way Analogues Simply Older than Blue?}
\label{sec:age}

\begin{figure}
    \includegraphics[width=85mm]{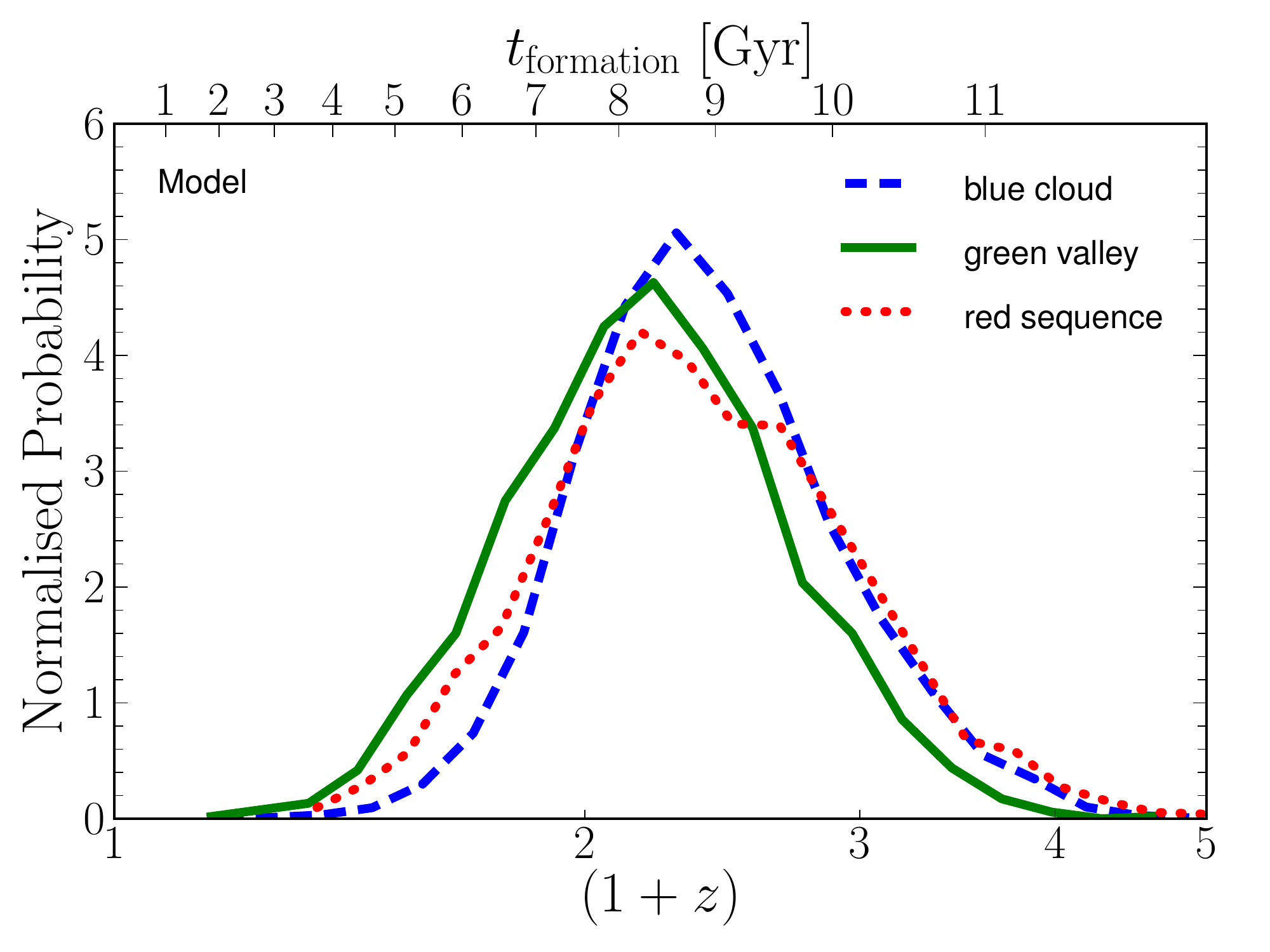}
    \caption{Normalized probability distribution of formation times for each color sub-sample of the model Milky Way/M31 analogues.  This time corresponds to the last snapshot in the simulation at which a galaxy possessed half of its $z$=0 stellar mass.  All three color sub-samples have similar formation time distributions indicating that age has little effect in determining the $z$=0 color distributions of our analogue galaxies. See the electronic edition of the Journal for a color version of this figure. }
    \label{fig:stellar_age}
\end{figure}

It seems reasonable to postulate that red sequence members are simply older than their blue counterparts, and so have progressed further in their color evolution.  In this scenario, green valley members would simply represent an intermediate population in terms of age.  

Figure \ref{fig:stellar_age} shows the distribution of formation times for red, green and blue model Milky Way/M31 analogues, defined to be the time when the galaxy had reached half its final stellar mass \citep[]{Croton2007}. Immediately obvious is the similarity in the formation time distributions of all three color sub-samples.  Qualitatively similar results are found when we define formation time to be $10\%$ or $90\%$ of the final stellar mass, or use halo mass instead. This suggests that red or green Milky Way-type galaxies are not simply older versions of their blue counterparts.  Based on the histories of our model galaxies, we can hence discount formation time as being a critical property on which spiral galaxy color depends.

\subsection{Rapid or Gradual Shut-down of Milky Way Star Formation?}
\label{sec:shut-down-rate}

The fractional abundance of Galaxy Zoo Milky Way-like green valley or red sequence galaxies is approximately 15\%.  For those model analogues which have crossed on to the red sequence in the last $\sim 10$ Gyr, we find the average time spent in the green valley to be $\sim1.5$ Gyr.  

The time-scale for passive evolution across the green valley after an instantaneous shut down of star formation is typically around a Gyr, somewhat shorter than the transit time found in the model (as measured from the same start and end color points). This suggests that it is unlikely that star formation is simply `switched off', but instead undergoes a somewhat more gradual decline \citep[]{Baldry2004Proc,Balogh2010}.  This is also in qualitative agreement with \citet[]{Masters2010b} who similarly suggest a gradual shutdown of star formation associated with their sample of red passive spirals drawn from Galaxy Zoo.

\subsection{Cold Gas Depletion in Spiral Galaxies?}
\label{sec:curtailing-sf}

\begin{figure}
    \includegraphics[width=85mm]{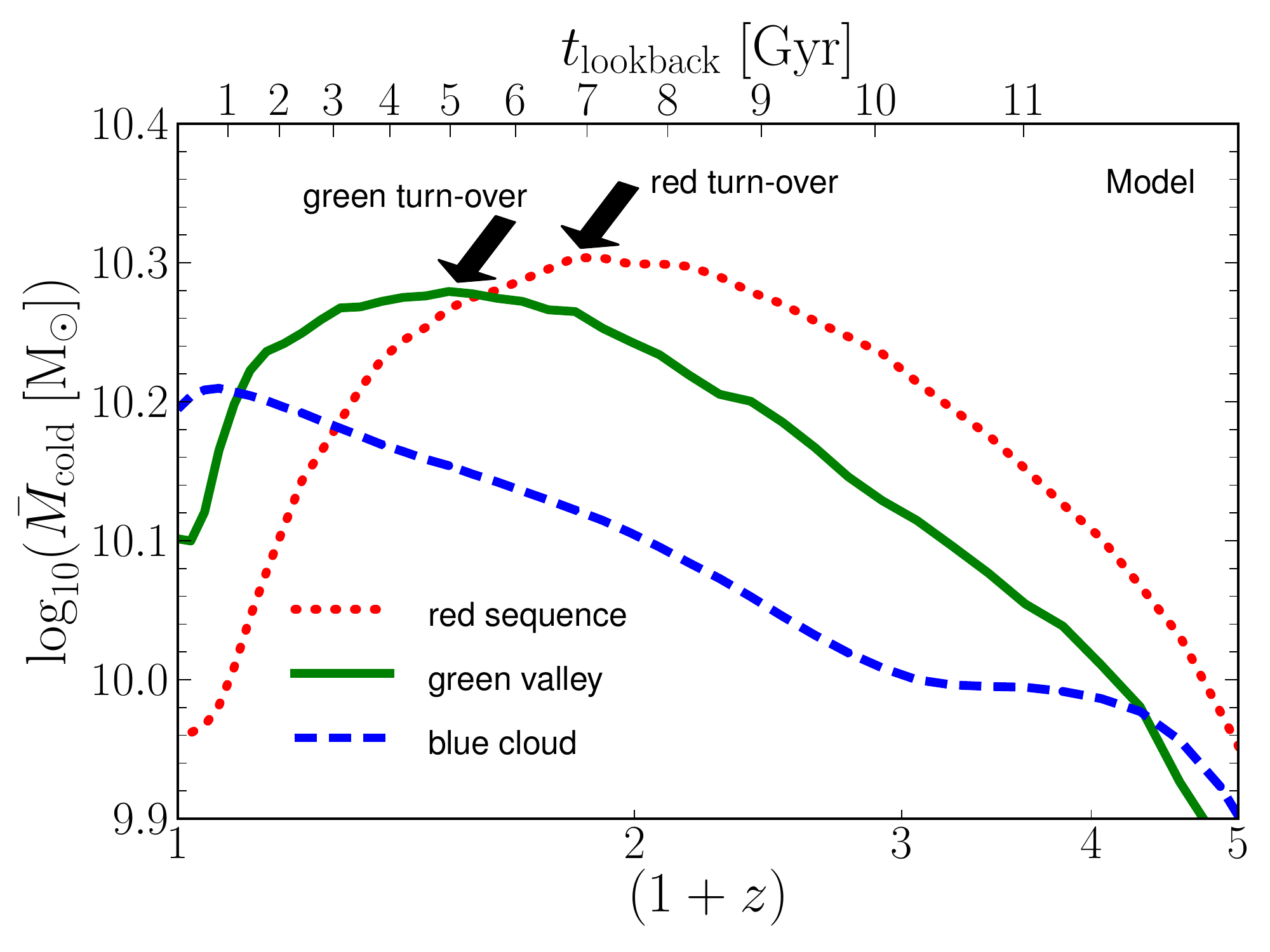}
    \caption{\label{fig:cold-mass} Plot of the mean cold gas mass of each color sub-sample of our model Milky Way/M31 analogues as a function of look-back time.  There is an obvious trend for red galaxies to have less cold gas than their bluer counterparts at $z$=0.  All color samples have an approximately equal peak mean cold gas mass, with the difference at $z$=0 being due to the time at which the mean cold gas content began its decline (as indicated by the arrows). See the electronic edition of the Journal for a color version of this figure.}
\end{figure}

Due to their typically quiescent merger histories, it is more likely that in-situ processes, rather than merger related processes, are responsible for stifling star formation in Milky Way-like galaxies. In-situ star formation in the model is directly driven by the amount of cold gas in, and the dynamical time-scale of, the galactic disk \citep[]{Kennicutt1998}. Simply put, the absence of cold gas implies the absence of star formation. Hence, we focus here on the evolving cold gas content of Milky Way-like galaxies in our model. 

Cold gas in galaxies is regulated by five primary mechanisms: cooling flows from the halo, star formation, supernova feedback, AGN heating, and galaxy-galaxy mergers/interactions. The complex interplay between each is non-trivial to predict \textit{a priori}, requiring the use of models and simulations. 

Figure~\ref{fig:cold-mass} shows the mean cold gas mass for red, green and blue model Milky Way-like galaxies as a function of look-back time. The evolution of cold gas shows a consistent trend across all three color bins: a growth in cold gas content from high redshift which peaks at approximately the same mass, followed by a decline at lower redshift as the galaxy moves off the blue cloud. Importantly, this turn-over occurs at earlier times for redder galaxies (black arrows). Notice the clear deficit of cold gas associated with local red spirals - a factor of two lower than blue spirals on average.

In general, cooling flows are the dominant source of fresh cold gas in the Milky Way/M31 analogue sample in our model.  Any mechanism which leads to the net reduction of cold gas mass will thus need to remove the gas at a faster rate than it is being replenished from the surrounding halo as the system grows with time.

\subsection{Cold Gas Heating in Spiral Galaxies?}
\label{sec:radio-mode-heating}

We now identify the three primary mechanisms thought to be responsible for reducing the abundance of disk gas in massive, gravitationally dominant spiral galaxies, and discuss each of these in turn.

First is the depletion of gas due to galaxy-galaxy mergers, which rapidly transforms disk gas into stars via starbursts. We have already argued that major mergers are not prevalent in our sample of Sb/c galaxies (by selection) and hence do not play a significant role here.  We also find that minor merger bursts only alter the gas mass, on average, by at most 5\% at $z=0$. The trend seen in Figure~\ref{fig:cold-mass} is therefore not due to mergers.

Second is the heating associated with supernova feedback following episodes of star formation. Supernova feedback acts to remove gas from the disk and return it to the surrounding halo medium. Supernova are certainly acting in our model in galaxies in this mass range. However, the energy available to remove gas from an $L^*$ galaxy disk is significantly smaller, on average, than that needed to offset the cooling rate from an $L^*$ host halo. In addition, star formation is needed to produce supernova and hence supernova feedback is not a plausible path to a population of long-term star formation deficient (i.e. red) galaxies.

\begin{figure}
    \includegraphics[width=85mm]{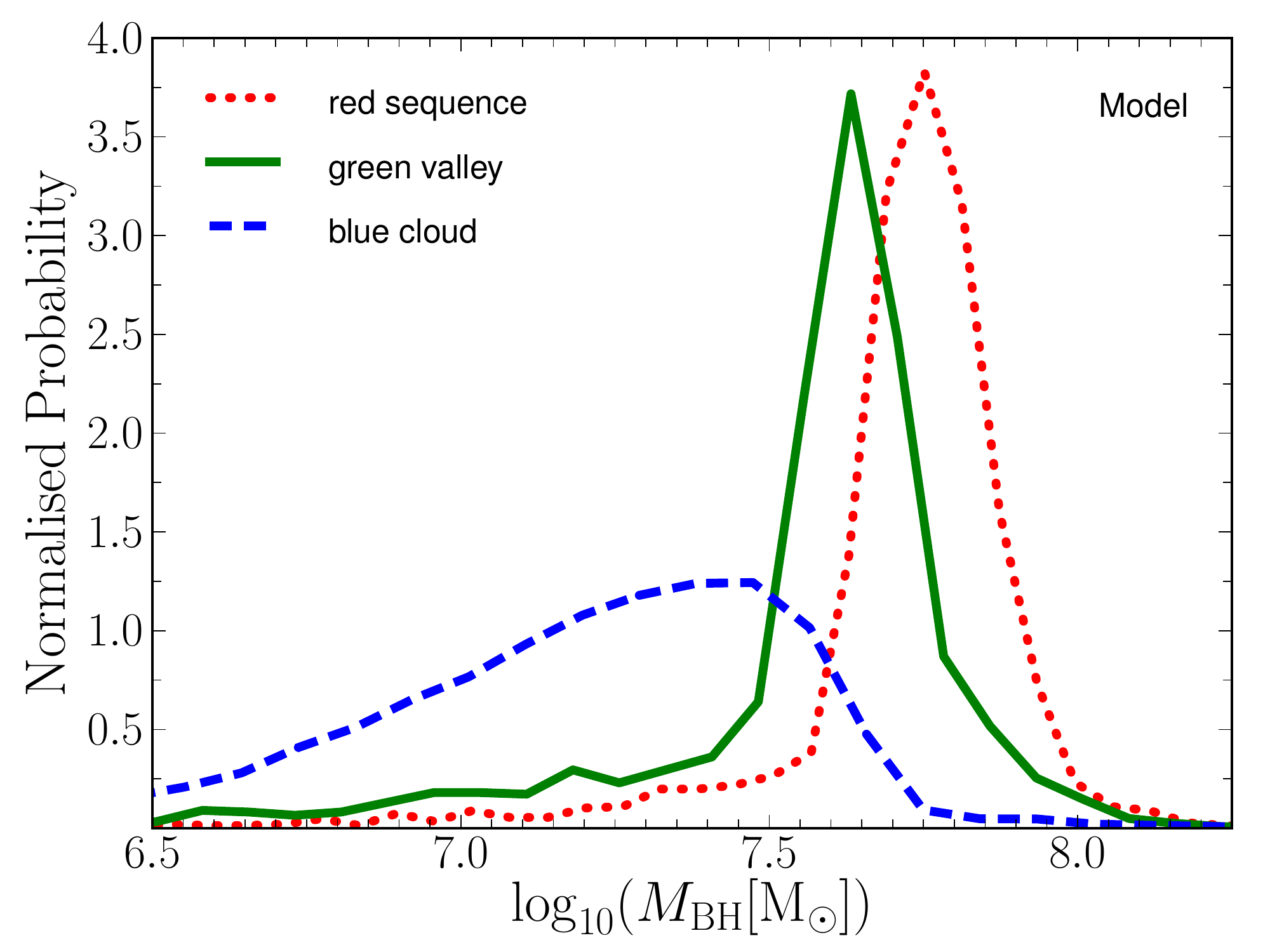}
    \caption{\label{fig:BH_mass} Normalized probability distribution of black hole masses for red, green and blue galaxies drawn from our model Milky Way/M31 analogue sample.  Red galaxies are, on average, a factor of two more massive than blue counterparts.  There is also a notable trend for the spread in black hole masses to decrease for redder galaxies. See the electronic edition of the Journal for a color version of this figure.\vspace{0.1em}}
\end{figure}

The third mechanism to reduce the abundance of disk gas, and hence star formation, is some form of active black hole heating. Black holes provide a plausible way of regulating the supply of halo gas into the disk through periodic heating cycles of low luminosity feedback \citep{Croton2006}. Their inclusion in galaxy formation models has provided a solution to a number of longstanding problems in galaxy formation theory, such as the overcooling problem in cluster halos, and the turn-over at the bright end of the galaxy luminosity and stellar mass functions.

Figure~\ref{fig:BH_mass} shows the distribution of black hole masses in our model analogue galaxies. As found in several works \citep[e.g.][]{Croton2006, Croton2007}, such masses are consistent with the observed local $M_{\rm BH}$--$M_{\rm bulge}$ relation. Even so, Figure~\ref{fig:BH_mass} reveals that red sequence Milky Way/M31 analogues tend to have a mean central black hole mass that is approximately two times larger than that of their blue cloud counterparts. When such galaxies become active, the heating is similarly larger by a comparable factor. In the model, this extra heating provides enough energy to suppress cooling and stifle the supply of fresh gas in a way that smaller holed spirals cannot\footnote{Remember that $L^*$ galaxies straddle an observed tipping point between star forming and passive, and the model is constrained to mimic this observation.}.

\subsection{Spiral Galaxies on the Edge}
\label{sec:spirals-on-edge}

AGN heating is the dominant physical process responsible for the decline in star formation rates in our model Milky Way/M31 analogue sample, and hence explains the green and red Sb/c model galaxy subpopulations.  There is some observational support for such a mechanism acting in real Milky Way-like objects.  For example, the fractional abundance of galaxies hosting active black holes is concentrated towards the green valley, supporting its theoretical use as a shut-down mechanism. \citet[]{Schawinski2010} suggest that the Milky Way \emph{should} lie in the highest duty cycle region of the color--stellar mass plane of late-type galaxies. Furthermore, \citet[]{Masters2010b} select a sample of face-on spirals with passive disks from Galaxy Zoo and find an increased fraction of Seyfert+LINER galaxies; a commonly suggested source of LINER emission in galaxies is low-luminosity AGN.  Finally, several studies \citep[e.g.,]{Giordano2010,Masters2011} find an increased prevalence of optical bars in redder spiral galaxies, with \citet[]{Masters2010b} also noting an increased bar fraction in their face-on, passive sample. Such bars have long been favored as a mechanism to channel gas onto a black hole through instabilities, and may be relevant to our understanding of the Milky Way, which is purported to have a significant bar \citep[]{Blitz1991}.

Tentative support also exists to suggest that Sgr A$^*$ (the central black hole of the Milky Way) itself, may have been considerably more active than its current state in the recent past.  For example, the observed luminosity of the Milky Way's black hole may have been almost 5 orders-of-magnitude brighter as little as 100 years ago \citep[]{Terrier2010}, and there have been recent claims of observed gamma ray bubbles seen above and below the galactic plane, supposedly as a result of an energetic event in the Galaxy's central region dated to within the past 10 Myr \citep[]{Su2010}.  Recent research has suggested that this energetic event is most likely due to past AGN activity \citep{Guo2011}.  Still, we stress that we are not aware of any definitive evidence to suggest that either the Milky Way or M31 would be classified as active galaxies if observed from a more privileged position in time or space.  If it proves that no recent AGN activity has indeed occurred in these two galaxies then for AGN to remain as a viable mechanism responsible for causing their reddening, a large time delay between any AGN event and the associated suppression of star formation would be required.

\begin{figure*}
   \begin{center}
   \begin{minipage}{160mm}
   \includegraphics[width=160mm]{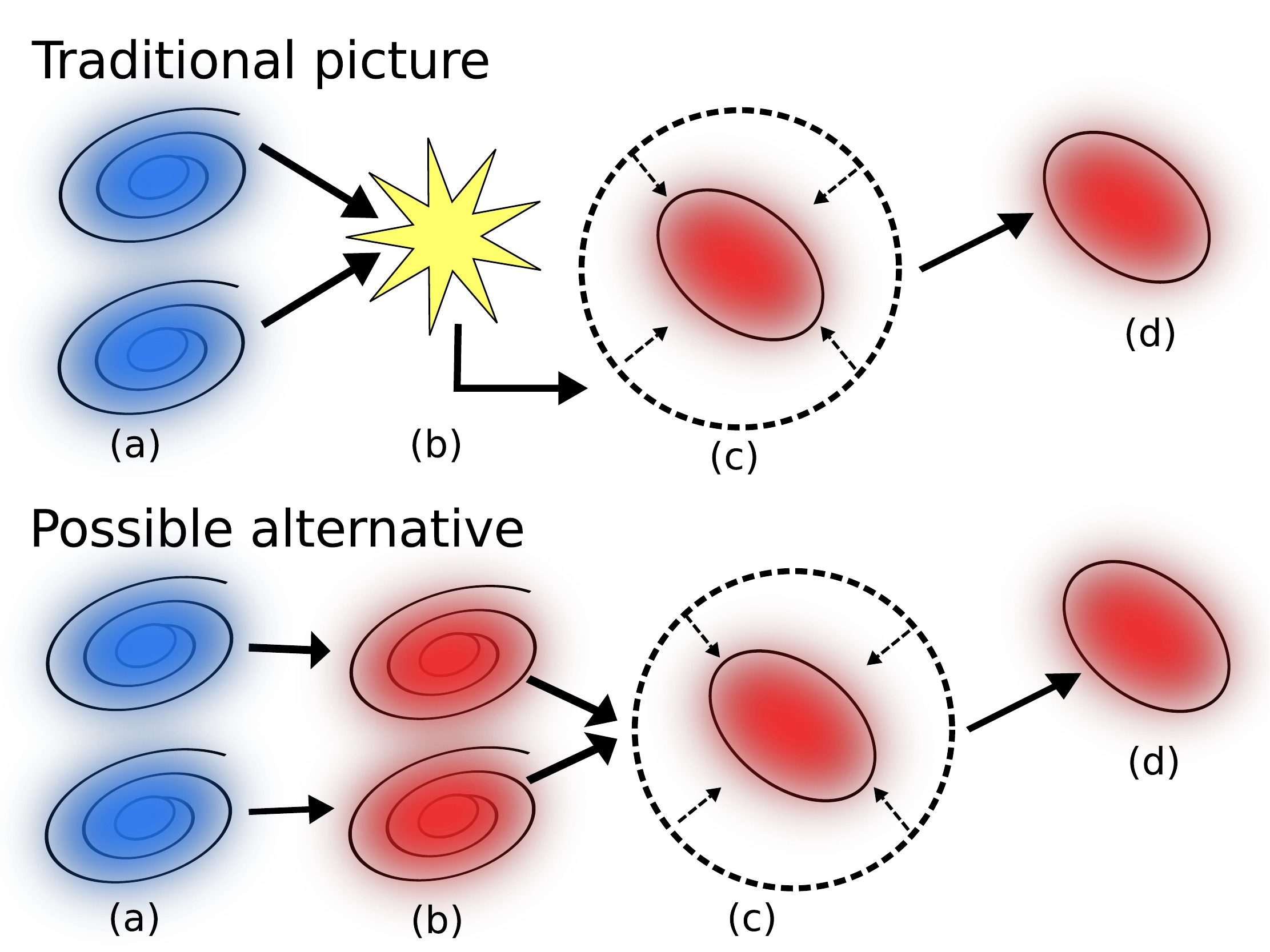}
    \caption{\label{fig:form-scheme} [\emph{Upper}] Illustration of the standard paradigm for galaxy evolution in a hierarchical universe. [\emph{Lower}] The alternative evolution suggested by a green, passively evolving population of spiral galaxies.  The main difference is the lack of a quasar mode phase in the alternative scenario. This is due to the depletion of cold gas, and the associated reddening of the stellar populations, which has already occurred in both galaxies before the merger event. With not enough cold gas to act as fuel, no powerful quasar mode burst occurs.  See the \S\ref{sec:spirals-on-edge} for a detailed discussion. See the electronic edition of the Journal for a color version of this figure.}
\end{minipage}
\end{center}
\end{figure*}

One thing we can say for certain is that $L^*$ galaxies, such as the Milky Way, occupy a unique position in the hierarchy of galaxy evolution that borders both star forming and passive galaxies on average. Their position at the `knee' of the galaxy luminosity function corresponds to the point where star formation quenching starts to become efficient \citep[]{Benson2003} and also to the transition valley region of the $D_n(4000)$--stellar mass plane (see Figure~1 of \citealt[]{Kauffmann2003}).  Thus, the significance of the separation in black hole mass between red and blue spirals seen in Figure~\ref{fig:BH_mass} appears to simply be a consequence of our model Milky Way/M31 analogues straddling this observationally required tipping point; the more massive holed spirals are forced onto the red sequence (in our model through quenching via AGN heating) and the less massive holed spirals remain fixed in the blue cloud. Our model green valley Milky Way-like galaxies may simply represent the mid-point of such evolution.

Although AGN heating is the physical mechanism responsible for truncating star formation in green and red Sb/c galaxies in the model, we note that we currently have few clear ways to distinguish this quenching mechanism from others that share similar scaling relations (at least given the current observational data).  The form of the invoked heating is something which has primarily been motivated by galaxies in more massive halos (e.g. clusters), and it is natural to extend this down to L$^*$ objects as a first approximation. We are thus provided with a valuable opportunity for both theorists and observers. Studies of a wider range of physically motivated quenching mechanisms, especially in lower mass L$^*$-scale dark matter halos, should allow modelers to find consistency with a wider range of observed galaxy populations. In harmony with this, programs and upcoming survey instruments (such as GAIA and HERMES), will allow observers to accurately quantify the nature of the color and SFR transformation that our own galaxy may be currently undergoing.

No matter what the cause, if the Milky Way and M31 truly are green valley galaxies and are transitioning on to the red sequence (perhaps slowly), as suggested here by their global colors and star formation rates, then this suggests a new paradigm of massive galaxy death which we outline in Figure~\ref{fig:form-scheme} and which is somewhat different to the conventional picture.  In the traditional, hierarchical paradigm of galaxy evolution (upper), a galaxy's life on the blue cloud \emph{(a)} ends in a gas rich major merger with another galaxy. This triggers the rapid accretion of cold gas on to the central black hole, seen observationally as a quasar \emph{(b)}. The quasar rapidly fades (over a few 100 Myr) to much lower luminosities \citep[]{Hopkins2006}, and heating from this low luminosity AGN then suppresses the further cooling of gas onto the galaxy over much longer timescales \emph{(c)}. This limits both the reformation of a new galactic disk, locking in the spheroidal morphology that resulted from the initial merger, and future star formation. The galaxy finally retires onto the red sequence as a passive `red and dead' elliptical \emph{(d)}. It may still grow through subsequent dissipationless `dry' mergers, but these only serve to maintain its color and morphology \citep[]{Faber2007}. 

Our own Local Group galaxies paint an alternative picture (lower panel of Figure~\ref{fig:form-scheme}). In this case, both the Milky Way and M31 are expected to merge in the next $\sim 5$ Gyr \citep[]{Cox2008}. If our current understanding of the transit time across the green valley is correct (a few Gyr), and if both galaxies are not `special' long-term residents of this region, then they should have already quiescently evolved on to the red sequence before the merger event. The eventual collision will thus be gas poor and it is likely that no quasar of significance will occur. The fading disks will still be destroyed, again resulting in a remnant elliptical galaxy with a mass approximately $2 L^*$. Subsequent evolution through dry mergers should then re-establish the standard paradigm.

This alternative picture is supported by a number of other observations of external galaxies.  For example, \citet[]{Darg2010} found a small merger fraction of $\sim$5\% in the local universe, thus suggesting that mergers may be less important to current galaxy evolution than previously considered.  Also it has been demonstrated that in the Galaxy Zoo sample, the rate of change of galaxy color from blue to red with environmental density is faster than the associated rate of change from spiral to elliptical \citep[]{Bamford2009,Skibba2009}.

\section{Conclusions}
\label{sec:conclusions}

Using the current best estimates of stellar masses, color indices and star formation rates for the Milky Way and M31, we investigate the current evolutionary position of these two galaxies through a comparison with an observational sample of galaxies drawn from SDSS / Galaxy Zoo1 data \citep[]{Bamford2009}.  Our key observational results can be summarized as follows:

\begin{itemize}
\item M31 is broadly consistent with being a member of the green valley population of the color vs. stellar mass diagram.  The large uncertainty in the global color of the Milky Way makes a similar conclusion regarding its color classification - based solely on this quantity - impossible. (\S\ref{sec:obs-results}).
\item At least 1 in 6 Milky Way-type galaxies in the local Universe lie red-ward of the blue cloud (assuming our most conservative selection criteria including galaxy orientation and Hubble classification) (\S\ref{sec:model-results}).
\end{itemize}

We then utilize the semi-analytic galaxy formation model of \citet[]{Croton2006} to generate a sample of theoretical Milky Way/M31 analogues.  After showing that this model is able to qualitatively reproduce the observed bimodality in the color vs. stellar mass plane for Milky Way-type galaxies, we use it to support our findings and elucidate the dominant physical mechanisms which may be at play:

\begin{itemize} 
\item The observed star formation rates of the Milky Way and M31 are consistent with both galaxies being categorized as green valley members when compared to the distribution of model Milky Way/M31 analogue star formation rates in different color bins (\S\ref{sec:model-results}).
\item There is no significant correlation of formation times with color in our model Milky Way/M31 analogue galaxies, suggesting that the red sequence analogues are not simply older versions of their blue counterparts (\S\ref{sec:age}).  Also, red model Milky Way/M31 galaxies transition across the green valley somewhat slower than what would be expected if star formation was instantaneously truncated (\S\ref{sec:shut-down-rate}).
\item The reduction of the availability of cold gas is the main cause for the decline of star formation in green and red model analogue galaxies. In the model this is a result of heating from active supermassive black holes. Green model Milky Way/M31 analogues straddle the tipping point between efficient and inefficient (or no) AGN feedback (\S\ref{sec:curtailing-sf},\S\ref{sec:spirals-on-edge}).
\item As the Milky Way and M31 are generally not considered to be active galaxies, our results suggest that alternative processes may be acting (although their precise nature is not currently obvious), to explain the position of these two galaxies in the color--stellar mass plane.  If low level AGN activity is responsible then it would require the presence of a considerable time delay between the feedback event and the truncation of star formation in such galaxies (\S\ref{sec:spirals-on-edge}).
\end{itemize}

Our key result, that both the Milky Way and M31 are possibly `green', has a number of important implications.  Although these galaxies are by no means atypical, they may not be examples of archetypal star forming spirals.  Instead, they possibly represent a population of galaxies in the midst of a transitional process.  Recent observations reveal that as many as 60\% of all galaxies which transition on to the red sequence have undergone a quiescent disk phase \citep[]{Bundy2010} and red sequence spiral galaxies are not uncommon \citep[e.g.]{Masters2010b}.  Care must therefore be exercised when using the Milky Way as a benchmark in studies of external galaxies.

The standard paradigm of galaxy evolution is for merger events to drive major changes in star formation and hence the bulk movement of objects on the color--stellar mass diagram.  The presence of high stellar mass spiral galaxies in the green valley and red sequence regions is somewhat contrary to this picture (Figure~\ref{fig:form-scheme}).  The fact that both our own galaxy and our nearest neighbor may be actively taking part in this alternative evolutionary scenario reveals their possible eventual fate.  The consequences for our understanding of galaxy evolution are significant \citep[]{Peebles2010} and serve to highlight the wide-reaching importance of Galactic experiments such as HERMES and GAIA, as well as APOGEE \citep[]{AllendePrieto2008} and upcoming cold gas surveys with ASKAP, ALMA \citep[]{Johnston2008,Minniti2010}. The Milky Way may provide us with a rare opportunity to examine, in exquisite detail, a galaxy actively experiencing one of the most important global transformations in galactic evolution.

\bibliography{Mutch_Refs}

\acknowledgments

SM is supported by a Swinburne University SUPRA postgraduate scholarship. DC acknowledges receipt of a QEII Fellowship awarded by the Australian government. GP is supported by the ARC DP program.  The authors would like to thank S. Bamford and the Galaxy Zoo team for kindly supplying the observational data used in this work.  The visual classification of the Galaxy Zoo galaxies was made by more than 100,000 volunteers.  Their contributions are acknowledged at http://www.galaxyzoo.org/Volunteers.aspx.  The authors would also like to thank C. Flynn, L. Spitler and the anonymous referee for their extremely helpful comments and suggestions.  The Millennium Simulation used in this paper was carried out by the Virgo Supercomputing Consortium at the Computing Centre of the Max-Planck Society. Semi-analytic galaxy catalogues from the simulation are publicly available at http://www.mpa-garching.mpg.de/millennium/.

\end{document}